\documentclass[letterpaper, 12 pt, conference, onecolumn]{ieeeconf}  

\IEEEoverridecommandlockouts                              

\usepackage[pdftex, pdfstartview={FitV}, pdfpagelayout={TwoColumnLeft},bookmarksopen=true,plainpages = false, colorlinks=true, linkcolor=black, citecolor = black, urlcolor = black,filecolor=black , pagebackref=false,hypertexnames=false, plainpages=false, pdfpagelabels ]{hyperref}

\usepackage[authormarkuptext=name,addedmarkup=bf,authormarkupposition=right]{changes}
\definechangesauthor[name={J.~H.}, color={red}]{jh}
\definechangesauthor[name={B.~L.}, color={blue}]{bl}

\makeatletter
\def\set@curr@file#1{%
  \begingroup
    \escapechar\m@ne
    \xdef\@curr@file{\expandafter\string\csname #1\endcsname}%
  \endgroup
}
\def\quote@name#1{"\quote@@name#1\@gobble""}
\def\quote@@name#1"{#1\quote@@name}
\def\unquote@name#1{\quote@@name#1\@gobble"}
\makeatother
\usepackage{graphicx}

\usepackage{epstopdf}
\usepackage{amssymb,amsmath}

\usepackage{amsthm}

\usepackage{subfig}
\usepackage[font=footnotesize]{caption}
\usepackage{xcolor}
\usepackage{units}
\usepackage{algorithm}
\usepackage[noend]{algpseudocode}
\usepackage{balance}
\usepackage[sort,compress]{cite}
\usepackage{tabularx}
\usepackage{nameref}

\newtheoremstyle{named}{}{}{\itshape}{}{\bfseries}{.}{.5em}{\thmname{#1}\thmnumber{ #2}\thmnote{ (#3)}}
\theoremstyle{named}
\newtheorem{theorem}{Theorem}
\theoremstyle{named}

\theoremstyle{named}
\newtheorem{proposition}{Proposition}
\theoremstyle{named}
\newtheorem{lemma}{Lemma}

\newtheoremstyle{namedDef}{}{}{}{}{\bfseries}{.}{.5em}{\thmname{#1}\thmnumber{ #2}\thmnote{ (#3)}}
\theoremstyle{namedDef}
\newtheorem{definition}{Definition}

\newtheoremstyle{myproblem}{}{}{}{}{\bfseries}{--}{.5em}{\thmname{#1}\thmnumber{ #2} \thmnote{#3}}
\theoremstyle{myproblem}

\theoremstyle{definition}
\newtheorem{assumption}{Assumption}

\theoremstyle{remark}
\newtheorem{remark}{Remark}

\usepackage[hang,flushmargin]{footmisc}

\usepackage[capitalize]{cleveref}
\crefformat{equation}{(#2#1#3)}
\Crefformat{equation}{Equation~(#2#1#3)}
\Crefname{equation}{Equation}{Eqs.}

\usepackage{bm}
\usepackage{bbm}

\usepackage{accents}

\title{\LARGE \bf
Contraction Metrics in Adaptive Nonlinear Control
}

\author{Brett T. Lopez$^{1}$ and Jean-Jacques E. Slotine$^{2}$
\thanks{$^{1}$NASA Jet Propulsion Laboratory, California Institute of Technology, Pasadena, CA, {\tt\small btlopez@mit.edu}}
\thanks{$^{2}$Nonlinear Systems Laboratory, Massachusetts Institute of Technology, Cambridge, MA, {\tt\small jjs@mit.edu}}
}

\begin{document}

\maketitle
\thispagestyle{plain}
\pagestyle{plain}

\begin{abstract}
\noindent Lyapunov stability theory is the bedrock of direct adaptive control.
Fundamentally, Lyapunov stability requires constructing a distance-like function which must decrease with time to ensure stability.
Feedback linearization, backstepping, and sum-of-squares optimization are common approaches for constructing such a distance function, but require the system to possess certain inherent/structural properties or involves solving a non-convex optimization problem.
These restrictions/complexities arise because Lyapunov stability theory relies on constructing an \emph{explicit} distance function.
This work uses contraction metrics to derive an adaptive controller for stabilizable nonlinear systems by constructing a distance-like function \emph{differentially} rather than explicitly.
Because stabilizability is  in fact equivalent to the existence of a contraction metric, the proposed approach is significantly more general than available  results in the literature.
In particular, the method can be applied to underactuated systems.
More broadly, it can also be used in transfer learning where a feedback controller has been carefully learned for a nominal system, but needs to remain effective in the presence of significant but structured variations in parameters.
Simulation results illustrate the approach.
\end{abstract}

\section{Introduction}

Adaptive control has long been considered an effective method for stabilizing uncertain dynamical systems.
A number of approaches have thus been developed for nonlinear systems, including techniques based on feedback linearization \cite{taylor1988adaptive,kanellakopoulos1989robustness}, sliding mode or boundary layer control \cite{ambrosino1984variable,slotine1986adaptive}, and backstepping \cite{kanellakopoulos1991systematic,krstic1992adaptive}. 
Many of the aforementioned techniques require the system possess certain inherent or structural properties.
In particular the system must be controllable and integrable for feedback linearization or in strict-feedback form for backstepping.
If parameter or disturbance bounds are known then sliding mode and its variants can be applied to achieve additional robustness so long as the system is feedback linearizable or in strict-feedback form.
Alternatively, depending on the characteristics of the uncertainty, one could instead construct a control Lyapunov function (CLF) directly.
Applying Sontag's universal formula \cite{sontag1989universal} then produces a nominal controller that can be augmented with adaptive feedback terms to stabilize the actual system.
If the dynamics are polynomial, sum-of-squares (SOS) optimization can be used to search for CLFs \cite{prajna2004nonlinear,tan2004searching}. 
While CLFs only require stabilizability, constructing CLFs is non-trivial since the set of solutions is not necessarily convex or connected \cite{rantzer2001dual}. 

Much of the early work in adaptive control for systems with parametric uncertainty leveraged the certainty equivalence principle to construct baseline controllers without considering the effects of adaptation.
Adaptive feedback terms are then derived using Lyapunov stability arguments for the actual system.
The certainty equivalence principle only holds when the uncertainty is in the span of the control input matrix; a criteria more formally known as the matching condition.
Relaxing the matching condition became the focus of the adaptive control community, leading to the extended matching condition (uncertainty one derivative away from the input) \cite{kanellakopoulos1989robustness} and eventually the more general unmatched condition via backstepping \cite{kanellakopoulos1991systematic,krstic1992adaptive}.
Adaptive CLFs were then introduced as a general framework for constructing adaptive controllers by casting the problem as a non-adaptive stabilization of an augmented system \cite{krstic1995control}.
A constructive procedure for adaptive CLFs based on backstepping was also proposed.
Backstepping is still the most effective method to stabilize systems with unmatched uncertainty. 

The above adaptive control strategies employs Lyapunov stability theory to prove the closed-loop system is stable.
An important but subtle byproduct of Lyapunov stability theory is the need to construct an \textit{explicit} distance function (or CLF) whose time derivative is negative definite.
Feedback linearization, backstepping, and SOS all rely on this notion of explicit distance functions to prove stability, albeit the methodology of doing so is quite different.
It is this dependency that restricts the use of feedback linearization and backstepping to specific types of systems, and one that leads to the computational challenges of searching directly for distance function with SOS.
Contraction analysis~\cite{lohmiller1998contraction} is an alternative to Lyapunov stability theory that does not require the explicit construction of a distance function, but rather employs local analysis of neighboring system trajectories to show convergence.
Intuitively, if any two arbitrary neighboring trajectories converge to each other then the system as a whole must converge to a nominal motion.
Global stability can then be inferred through local analysis of the infinitesimal displacement between trajectories, a property that has lead to several important results \cite{lohmiller1998contraction,wang2005partial,lohmiller2005contraction,wang2006contraction}.
As succinctly stated in~\cite{forni2013differential} about contraction
analysis, ``this approach brings differential geometry to the rescue of Lyapunov theory."
More recent work~\cite{manchester2017control} applies the notion of contraction to synthesizing controllers for nonlinear systems.
Through the use of so-called \emph{control contraction metrics}, \cite{manchester2017control} formalizes the idea that stabilizability of a system should be enough to design a controller, without requiring additional integrability and controllability conditions as in feedback linearization.
Further, the constructive conditions can be formulated as a linear matrix inequality, circumventing the computational challenges of SOS CLFs.
Contraction metrics have been successfully used in distributed economic MPC \cite{wang2017distributed}, tube MPC \cite{singh2017robust}, and learning stable dynamics \cite{blocher2017learning,singh2018learning}.

\textbf{Statement of contributions:}  
This work develops a direct adaptive control approach for stabilizable nonlinear systems with parametric matched or extended matched uncertainty using the control contraction metric framework \cite{manchester2017control,singh2017robust}. 
Since stabilizability is in fact \emph{equivalent} to the existence of a contraction metric \cite{manchester2017control,singh2018learning} the proposed approach is more general than what exists in the literature.
Common modifications to improve tracking error transients and robustness to disturbances or sensor noise, such as a deadzone or incorporating parameter bounds, can be immediately added to the proposed controller.
The generality of the approach implies that it can be combined with \textit{any} robust feedback policy, learned or otherwise, so long as a closed-loop contraction metric or explicit control Lyapunov function is known.
This includes underactuated systems where systematic synthesis of adaptive controllers is especially challenging \cite{pucci2015collocated,moore2014adaptive}.
Simulation results illustrate the effectiveness of the approach, using a non-invertible nonlinear system that cannot be put into strict-feedback form with matched or extended matched uncertainty.

\section{Problem Formulation \& Preliminaries}

Consider the nonlinear system 
\begin{equation}
\dot{x} = f(x) - \Delta(x)^\top \theta + B(x)u,
\label{eq:genDyn}
\end{equation}
with unknown parameters $\theta \in \mathbb{R}^p$ with dynamics $\Delta \in \mathbb{R}^{p\times n}$, state $x \in \mathbb{R}^n$, control input $u \in \mathbb{R}^m$, nominal dynamics $f$, and control input matrix $B$ with columns $b_i$ for $i=1,\dots,m$. 
The uncertainty satisfies the \textit{matching condition} if $\Delta(x)^\top \theta \in \text{span}\{B\}$ or the \textit{extended matching condition} if $\Delta(x)^\top \theta \in \text{span}\{B,~ad_fB\}$ where $ad_fB = \left[f,~B\right] = \frac{\partial f}{\partial x} B - \frac{\partial B}{\partial x} f$ is shorthand for the Lie bracket.
The goal of this work is to construct a feedback policy $u = \kappa(x,\hat{\theta})$ and adaptation law $\dot{\hat{\theta}} = \tau(x,\hat{\theta})$ for stabilizable nonlinear systems such that the closed-loop system \cref{eq:genDyn} is stable.

Differential geometry is central to contraction metrics so a brief review of key concepts is presented here. Let $\mathcal{M}$ be a smooth manifold (which will be $\mathbb{R}^n$ for this work) equipped with a Riemannian metric $M(x,t)$ that defines an inner product $\left< \cdot, \cdot \right>_x$ on the tangent space $T_x \mathcal{M}$ at every point $x$.
The Riemannian metric $M(x,t)$ defines local geometric notions such as angles, length, and orthogonality.
The directional derivative of a metric $M(x,t)$ along vector $v$ is expressed as $\partial_v M = \sum_i\frac{\partial M}{\partial x_i} v_i$.
Let $c: \left[0,1\right] \rightarrow \mathcal{M}$ be a regular (i.e., $\frac{\partial c}{\partial s} = c_s \neq 0$ $\forall s \in \left[0~1\right]$) parameterized differentiable curve.
The length $L$ and energy $E$ of curve $c$ are given by
\begin{equation*}
L(c,t) = \int \limits_0^1 \sqrt{c_s^\top M(c,t) c_s} ds \quad 
E(c,t) = \int \limits_0^1 c_s^\top M(c,t) c_s ds.
\end{equation*}
Let $\Xi(p,q)$ denote the family of curves connecting two points $p$ and $q$ such that $c(0) = p$ and $c(1) = q$.
The Riemannian distance between $p$ and $q$ is given by
\begin{equation*}
d(p,q,t) = \underset{c(s) \in \Xi(p,q)}{\mathrm{inf}} L(c,t)
\end{equation*}
where $E(p,q,t) = d(p,q,t)^2$. 
By the Hopf-Rinow theorem, under suitable conditions a minimizing curve known as a \textit{minimal geodesic} $\gamma: \left[0~1\right] \rightarrow \mathcal{M}$ is guaranteed to exist with the unique property $E(\gamma,t) = L(\gamma,t)^2 \leq L(c,t)^2 \leq E(c,t)$. 
The first variation of energy with respect to time is given by \cite{carmo1992riemannian}
\begin{equation}
\frac{1}{2}\dot{E}(c,t) =  \frac{\partial E}{\partial t} + \left.\left< c_s(s),~ \dot{c}(s)\right>\right|^{s=1}_{s=0} - \int \limits_0^1  \left< \frac{D}{\partial s}  c_s,~ \dot{c}\right> ds, 
\label{eq:dEdt_c}
\end{equation} 
where $\frac{D(\cdot)}{\partial s}$ is the covariant derivative.
For a minimizing geodesic $\gamma(s)$, $\frac{D\gamma_s(s)}{\partial s}=0$ so \cref{eq:dEdt_c} becomes
\begin{equation*}
\frac{1}{2}\dot{E}(\gamma,t) =  \frac{\partial E}{\partial t} + \left.\left< \gamma_s(s),~ \dot{\gamma}(s)\right>\right|^{s=1}_{s=0}.
\end{equation*}
The time argument in the Riemannian metric and Riemannian energy is dropped in the sequel for clarity.

\section{Review of Contraction Metrics}
Consider the nominal system
\begin{equation}
    \dot{x} = f(x) + B(x)u,
    \label{eq:nomDyn}
\end{equation}
and corresponding differential dynamics
\begin{equation*}
\dot{\delta}_x  = A(x,u)\delta_x + B(x)\delta_u,
\end{equation*}
where $A(x,u) = \frac{\partial f}{\partial x} + \sum_{i=1}^m \frac{\partial b_i}{\partial x}u_i$.
Consider the function $\delta_V = \delta_x^\top M(x) \delta_x$, which can be viewed as the differential Riemannian energy at point $x$.
Differentiating and imposing that the differential Riemannian energy decreases exponentially with rate $\lambda$, one obtains
\begin{equation}
\dot{\delta}_V  = \delta_x^\top \left(A^\top M + MA + \dot{M} \right)\delta_x + 2 \delta_x^\top MB\delta_u \leq -2 \lambda \delta_x^\top M \delta_x,
\label{eq:dclf_dot}
\end{equation}
where $\dot{M}(x) = \frac{\partial M}{\partial t} + \sum_{i=1}^n \frac{\partial M}{\partial x_i} f_i(x)$. 
\begin{definition}[Manchester and Slotine \cite{manchester2017control}]
The system \cref{eq:nomDyn} is said to be \textit{universally exponentially stabilizable} if, for any feasible desired trajectory $x_d(t)$ and $u_d(t)$, a feedback controller can be constructed such that for any initial condition $x(0) \in \mathbb{R}^n$, a unique solution to \cref{eq:nomDyn} exists that satisfies 
\begin{equation*}
\|x(t)-x_d(t)\| \leq  R \|x(0)-x_d(0)\|e^{-\lambda t},
\end{equation*}
where $\lambda,~R > 0$ are the convergence rate and overshoot, respectively, independent of the initial conditions.
\end{definition}

\begin{theorem}[Manchester and Slotine \cite{manchester2017control}]
If there exists a uniformly bounded metric $M(x)$ such that the following implication is true
\begin{equation}
\delta_x^\top MB = 0 \implies \delta_x^\top\left(A^\top M + MA + \dot{M} + 2 \lambda M \right)\delta_x \leq 0,
\label{eq:ccm_impl}
\end{equation}
for all $\delta_x \neq 0,~ x,~ u$
then system \cref{eq:nomDyn} is universally exponentially stabilizable via continuous feedback defined almost everywhere, and everywhere in a neighborhood of the target trajectory
\label{thm:ccm}
\end{theorem}

\begin{remark}
The CCM condition given by \cref{eq:ccm_impl} can lead to complex feedback controllers because $A(x,u)$ is dependent on $u$.
If an additional condition is imposed then simpler controllers can be obtained. 
Specifically, one can enforce the columns of $B(x)$ (denoted as $b_i(x)$ for $i=1,\dots,m$) to form a Killing vector field for the metric $M(x)$.  
Hence, if $\delta_x^\top MB = 0$ then the metric $M(x)$ must satisfy
\begin{equation}
\begin{aligned}
&\delta_x^\top \left(\frac{\partial f}{\partial x}^\top M + M\frac{\partial f}{\partial x} + \dot{M} + 2 \lambda M \right)\delta_x \leq 0 \\
&\partial_{b_i}M + \frac{\partial b_i}{\partial x}^\top M + M \frac{\partial b_i}{\partial x} = 0, ~~ \text{for}~i=1,\dots,m ,
\end{aligned}
\label{eq:ccm_cond}
\end{equation}
where the first condition ensures the dynamics orthogonal to the input are contracting and the second condition forces each $b_i(x)$ to be a Killing vector field for $M(x)$.
\end{remark}

\begin{remark}
The intuition behind Theorem~\ref{thm:ccm} is that the system must be naturally contracting in directions orthogonal to the control input.
This can be interpreted as a stabilizability condition for the differential dynamics of system \cref{eq:nomDyn}.
\end{remark}

The CCM condition given by \cref{eq:ccm_impl} can be transformed into a convex constructive condition for the metric $M(x)$ by a change of variables.
Let $\eta = M(x) \delta_x$ and $W(x) = M(x)^{-1}$ (commonly referred to as the \textit{dual metric}), then \cref{eq:ccm_impl} can be expressed as
$\eta^\top\left( W A^\top + A W - \dot{W} + 2 \lambda W \right)\eta \leq 0$ whenever $\eta^\top B=0$.
One can let $\eta = B_{\perp}$ where the columns of $B_{\perp}$ span the null space of the input matrix $B$ (i.e., $B_{\perp}^\top B = 0$).
Then the CCM condition \cref{eq:ccm_impl} can be expressed as $
B_{\perp}^\top\left( W A^\top + A W - \dot{W} + 2 \lambda W \right)B_{\perp} \leq 0$ which is convex since the only unknown $W(x)$ appears linearly in the inequality.
The stronger conditions given in \cref{eq:ccm_cond} can also be cast as convex constructive conditions for $W(x)$
\begin{equation*}
\begin{aligned}
& B_{\perp}^\top \left(W \frac{\partial f}{\partial x}^\top  + \frac{\partial f}{\partial x} W - \dot{W} + 2 \lambda W \right)B_{\perp} \leq 0 \\
-&\partial_{b_i}W + W \frac{\partial b_i}{\partial x}^\top + \frac{\partial b_i}{\partial x} W = 0, ~~ \text{for}~i=1,\dots,m.
\end{aligned}
\label{eq:dual_ccm_cond}
\end{equation*}

The existence of a contraction metric $M(x)$ is necessary and sufficient for stabilizability via \cref{thm:ccm}.
What remains is constructing a feedback controller that achieves universal exponential stability.
\cite{manchester2017control} proposed the differential controller $\delta_u = -\frac{1}{2}\rho(x)B(x)^\top M(x)\delta_x$ where $\rho(x)$ is a scalar function found simultaneously with the dual the metric $W(x)$ by solving the LMI $ W A^\top + A W - \dot{W} + 2 \lambda W - \rho B B^\top \prec 0$.
As discussed in \cite{manchester2017control}, other differential controllers, such as the pointwise min-norm controller \cite{primbs2000receding}, can be synthesized by interpreting the Riemannian energy as a differential CLF.
In the more general case, for any suitable differential controller $\delta_u$ that satisfies \cref{eq:dclf_dot}, the final controller takes the form
\begin{equation*}
k(x,x_d) = u_d - \int \limits_0^1 \delta_u(\gamma(s)) ds,
\end{equation*}
where the path integral is along a minimal geodesic $\gamma(s)$ connecting the current $x$ and desired state $x_d$.

\section{Adaptive Control with Contraction Metrics}
\label{sec:accm}
\subsection{Overview}
This section presents the main results of this article.
A nonlinear adaptive controller with contraction metrics for the matched uncertainty condition is first derived.
A modification to exponentially bound the tracking error transients when parameter bounds are known is then presented.
The matched result is expanded to the extended matched uncertainty condition through the introduction of the parameter-dependent contraction metric.
Common techniques that improve the robustness of adaptive controllers, such as a deadzone or incorporating parameter bounds, can be immediately added to the proposed controller.
Finally, the offline and online computation of the proposed approach is discussed.

\subsection{Matched Uncertainty}
\label{sub:matched}
First consider a nonlinear systems with uncertainty that lies in the span of the input matrix.

\begin{assumption}
The parametric uncertainty satisfy the \emph{matching condition} $\Delta(x)^\top \theta \in \text{span}\{B\}$.
\label{assumption:match}
\end{assumption}

By Assumption~\ref{assumption:match}, the uncertainty can be expressed as $\Delta(x)^\top \theta = B(x)\varphi(x)^\top \theta$ so system \cref{eq:genDyn} can be rewritten as
\begin{equation}
    \dot{x} = f(x) + B(x)\left[ u - \varphi(x)^\top \theta \right].
    \label{eq:matchedDyn}
\end{equation}
The remainder of this subsection will reference system \cref{eq:matchedDyn}. 
\cref{lemma:matched} simplifies the construction of a contraction metric for systems with matched uncertainty.

\begin{lemma}
\label{lemma:matched}
If the uncertainty satisfies Assumption~\ref{assumption:match} and a contraction metric $M(x)$ satisfies the stronger CCM conditions for the \emph{nominal system}, then the same metric satisfies the stronger CCM conditions for the \emph{true system}.
\end{lemma}

\begin{proof}
See \nameref{sec:appendix}.
\end{proof}

\begin{assumption}
\label{assumption:ccmExists}
A uniformly bounded contraction metric $M(x)$ exists for the nominal dynamics \cref{eq:matchedDyn}.
\end{assumption}

\begin{theorem}
\label{thm:matched}
    For an uncertain nonlinear system that satisfies Assumption~\ref{assumption:match} and Assumption~\ref{assumption:ccmExists}, the following adaptive feedback controller renders the closed-loop system asymptotically stable
     \begin{equation}
     \label{eq:auccmMatched}
        \begin{aligned}
            u &= u_{ccm} + \varphi(x)^\top \hat{\theta} \\
            \dot{\hat{\theta}} &= -\Gamma \varphi(x) B (x)^\top M(\gamma(1))\gamma_s(1)
        \end{aligned}
    \end{equation}
    where $\hat{\theta}$ is the current parameter estimate, $u_{ccm}$ is the CCM controller for the nominal system, $\varphi(x)$ is the basis vector for unknown parameters $\theta$,  $\gamma(s)$ is a minimizing geodesic, and $\Gamma \in \mathbb{R}^{p\times p}$ is a diagonal matrix with all positive elements that governs the rate of adaptation.
\end{theorem}

\begin{proof}
    See \nameref{sec:appendix}.
\end{proof}

The tracking error converges asymptotically to zero with the adaptive controller given by \cref{eq:auccmMatched} but better performance in terms of guaranteed convergence rate can be achieved if the parameters are assumed to be bounded.

\begin{assumption}
    \label{assumption:paramBounds}
    Each parameter $\theta_i$ belongs to the closed convex set $\Upsilon_{\theta_i} = \left\{\theta_i~|~\theta^-_i \leq \theta_i \leq \theta^+_i\right\}$.
\end{assumption}

\cref{lemma:ccmRobust}, inspired by \cite{yao1997adaptive} \cite[p.~78]{miroslav1995nonlinear}, shows how the Riemannian energy, and hence tracking error, can be exponentially bounded by augmenting a baseline CCM controller with additional robustness terms.

\begin{lemma}
    \label{lemma:ccmRobust}
    For an uncertain nonlinear system that satisfies Assumption~\ref{assumption:match}, Assumption~\ref{assumption:ccmExists}, and Assumption~\ref{assumption:paramBounds}, the following feedback controller ensures the closed-loop error transients is exponentially bounded
    \begin{equation}
    \label{eq:uccmRobust}
        u_i = u_{ccm,i} + \varphi_i(x)^\top \hat{\theta} - \kappa  b_i(x)^\top M\left(\gamma(1)\right)\gamma_s(1) \left|\left| \varphi_i(x)\right|\right|^2 , ~~ i = 1,\dots,m
    \end{equation}
    where $u_{ccm}$ is the baseline controller for the nominal dynamics, $\varphi_i(x)$ is the $i^\text{th}$ column of $\varphi(x)$, $\hat{\theta}$ is an estimate of the unknown parameters (without adaptation this can be chosen to be the mean of the known bounds), $\kappa > 0$ is a design parameter, and $b_i(x)$ is defined as before.
    Furthermore, the Riemannian energy $E$ satisfies the inequality
    \begin{equation}
    \label{eq:E_bound}
        \dot{E} \leq - 2\lambda E + K \left|\left|\tilde{\theta}_{\infty}\right|\right|^2
    \end{equation}
    where $\tilde{\theta}_{\infty} \in \mathbb{R}^p$ with $\tilde{\theta}_{\infty,i} := ~\theta^+_{i} - \theta^-_{i}$ and $K:=\frac{m}{2 \kappa }$.
\end{lemma}

\begin{proof}
See \nameref{sec:appendix}
\end{proof}

Combining \cref{eq:auccmMatched} with \cref{eq:uccmRobust} results in closed-loop system that asymptotically converges to zero with exponentially bounded transients. 

\begin{theorem}
    \label{thm:accm_transient}
    For an uncertain nonlinear system that satisfies Assumption~\ref{assumption:match}, Assumption~\ref{assumption:ccmExists}, and Assumption~\ref{assumption:paramBounds}, the following adaptive feedback controller renders the closed-loop system asymptotically stable with exponentially bounded transients
    \begin{equation}
    \label{eq:auccmRobust}
    \begin{aligned}
            u_i &= u_{ccm,i} + \varphi_i(x)^\top \hat{\theta} - \kappa  b_i(x)^\top M\left(\gamma(1)\right)\gamma_s(1) \left|\left| \varphi_i(x)\right|\right|^2 , ~~ i = 1,\dots,m \\
            \dot{\hat{\theta}} &= -\Gamma \varphi(x) B (x)^\top M\left(\gamma(1)\right)\gamma_s(1)
        \end{aligned}
    \end{equation}
    where all quantities are defined as before.
\end{theorem}

\begin{proof}
See \nameref{sec:appendix}
\end{proof}

\begin{remark}
    The exponential bound on the Riemannian energy in \cref{eq:E_bound} can be easily converted to a bound on the tracking error
    \begin{equation*}
        \|x(t)-x_d(t)\| \leq  R \|x(0)-x_d(0)\|e^{-\lambda t} + R \sqrt{\frac{K}{2 \lambda}} \left|\left|\tilde{\theta}_{\infty} \right|\right|\left[1-e^{-2\lambda t}\right]^{\frac{1}{2}},
    \end{equation*}
    which, in the context of robust model predictive control, describes the evolution of the tube dynamics with structured model uncertainty.
    Tube MPC with contraction metrics was explored in \cite{singh2017robust} but structured model uncertainty was not considered.
    Future work will extend \cite{lopez2019adaptive} by developing an adaptive robust MPC framework for stabilizable nonlinear systems using adaptive control, set membership identification, and contraction metrics. 
\end{remark}

\subsection{Extended Matched Uncertainty}
\label{sub:exMatched}
\cref{thm:matched} presented an adaptive controller for stabilizable nonlinear systems with matched uncertainty.
While this is a common assumption in adaptive control, several real-world systems such as electrical motors or systems with actuator dynamics possess uncertainty on derivative away from the input.
This subsection presents an adaptive nonlinear controller for such systems.
\begin{assumption}
\label{assumption:exMatched}
    The parametric uncertainty satisfies the extended matching condition 
    \begin{equation*}
        \Delta(x)^\top \theta \in \text{span}\{ad_fB\},
    \end{equation*} 
    where $B$ and $ad_fB$ are linearly independent vectors.
\end{assumption}

\begin{remark}
    The linear independence requirement in Assumption~\ref{assumption:exMatched} is equivalent to the dynamics one derivative away from the input being controllable. 
    This condition is necessary so the adaptive controller can stabilize the uncertain dynamics.
\end{remark}

For clarity let $\varrho(x)^\top \theta = \Delta(x)^\top \theta$.
\cref{sub:matched} already addressed the matched case so lets only consider deriving an adaptive controller for the system
\begin{equation}
    \label{eq:exMatchedDyn}
    \dot{x} = f(x) - \varrho(x)^\top \theta + B(x) u,
\end{equation}
where $\varrho(x)^\top \theta \in \text{span}\{ad_fB\}$.

\begin{definition}
\label{def:pdCcm}
    A \textit{parameter-dependent (dual) contraction metric} is a uniformly bounded metric, positive definite in both arguments, that satisfies the (dual) CCM conditions for \textit{every} possible value of the unknown parameters $\theta$.
\end{definition}

\begin{remark}
    \cref{def:pdCcm} is analogous to the parameter-dependent Lyapunov function presented in \cite{gahinet1996affine,feron1996analysis} without the additional requirement of expressing the Lyapunov function explicitly.
\end{remark}

\begin{assumption}
    There exists a parameter-dependent contraction metric $M(x,\theta)$ that stabilizes the uncertain system. \label{assumption:exCCMexists}
\end{assumption}

In the context of adaptive control, \cref{def:pdCcm} is extremely useful since the unknown parameter $\theta$ can be replaced with the parameter estimate $\hat{\theta}$.
In doing so, however, the adaptation rate $\dot{\hat{\theta}}$ then appears in the derivative of the differential CLF
\begin{equation}
    \dot{\delta}_V  = \delta_x^\top \left(A^\top M + MA + \frac{\partial M}{\partial t} + \partial_f M \right)\delta_x + \sum_{i=1}^p\delta_x^\top \frac{\partial M}{\partial \hat{\theta}_i}\dot{\hat{\theta}}_i \delta_x + 2 \delta_x^\top MB\delta_u \leq -2 \lambda \delta_x^\top M \delta_x.
\label{eq:dclf_dot_adapt}
\end{equation}
The challenge of constructing adaptive controllers for unmatched uncertainty is now clear: there is no guarantee that the differential controller $\delta_u$ in \cref{eq:dclf_dot_adapt} can cancel the term $\sum_{i=1}^p\delta_x^\top \frac{\partial M}{\partial \hat{\theta}_i}\dot{\hat{\theta}}_i \delta_x$.
Furthermore, there are now two unknowns in \cref{eq:dclf_dot_adapt}, the contraction metric $M(x,\hat{\theta})$ and adaptation rate $\dot{\hat{\theta}}$.
Fortunately, for the special case where the uncertainty satisfies the extended matched condition, a parameter-dependent contraction metric can be synthesized independently of the adaptation law. 

\begin{lemma}
    If Assumption~\ref{assumption:exMatched} and Assumption~\ref{assumption:exCCMexists} are true then
    \begin{equation}
    \label{eq:ex_rate_cond}
        \delta_x^\top \frac{\partial M}{\partial \hat{\theta}_i} \delta_x = - 2 \delta_x^\top M b_k \left[ \frac{\partial \varrho_i}{\partial x_1} \cdots 0 \right] \delta_x
    \end{equation}
    for each $\hat{\theta}_i$ where $\varrho_i(x)$ is the $i^{th}$ row vector of $\varrho(x)$ and $b_k(x)$ is the column vector of $B(x)$ such that $\varrho_i(x)^\top \theta_i \in \mathrm{span}\{ad_f b_k\}$.
    Furthermore, the contraction metric does not depend on the adaptation law.
    \label{lemma:exMatched}
\end{lemma}

\begin{proof}
    See \nameref{sec:appendix}.
\end{proof}

\begin{remark}
    \cref{lemma:exMatched} is a generalization of the result in \cite[p.~126]{miroslav1995nonlinear} since it utilizes a differential CLF, not an explicit one.
\end{remark}

\begin{remark}
    The zero element of the bracketed term in \cref{eq:ex_rate_cond} is a consequence of the extended matched condition.
\end{remark}

Leveraging \cref{lemma:exMatched} and Assumption~\ref{assumption:exCCMexists}, a stabilizing adaptive controller for the extended matched uncertainty case can be derived.

\begin{theorem}
\label{thm:exMatched}
For an uncertain nonlinear system that satisfies Assumption~\ref{assumption:exMatched} and Assumption~\ref{assumption:exCCMexists}, the following adaptive feedback controller renders the closed-loop system asymptotically stable
\begin{equation}
    \begin{aligned}
        u &= u_{ccm} +  \mathbbm{1} \sum_{i=1}^p \dot{\hat{\theta}}_i \int \limits_0^1 \left[ \frac{\partial \varrho_i\left(\gamma(s)\right)}{\partial x_1} \cdots 0 \right] \gamma_s(s)  ds \\
        \dot{\hat{\theta}} &= - \Gamma \varrho(x)  M\left(\gamma(1),\hat{\theta}\right)\gamma_s(1) 
    \end{aligned}
\end{equation}
where $B(x)\mathbbm{1} = b_k(x)$ so $\varrho_i(x)^\top \theta_i \in \text{span}\{ad_fb_k\}$,  $\hat{\theta}$ is the current parameter estimate, $u_{ccm}$ is the CCM controller for the current parameter estimate $\hat{\theta}$, $\varrho(x)$ is the basis vector with columns $\varrho_i(x)$ for the unknown parameter $\theta_i$, $\gamma(s)$ is a minimizing geodesic, and $\Gamma \in \mathbb{R}^{p\times p}$ is a diagonal matrix with all positive elements that governs the rate of adaptation.
\end{theorem}

\begin{proof}
See \nameref{sec:appendix}.
\end{proof}

\begin{remark}
    The controller and adaptation law derived for matched and extended uncertainty can be immediately combined into a single stabilizing adaptive controller
    \begin{equation}
    \label{eq:comb_controller}
        \begin{aligned}
            u &= u_{ccm} + \mathbbm{1} \sum_{i=1}^p \dot{\hat{\theta}}_{em,i} \int \limits_0^1 \left[ \frac{\partial \varrho_i\left(\gamma(s)\right)}{\partial x_1} \cdots 0 \right] \gamma_s(s)  ds + \varphi(x)^\top \hat{\theta}_m \\
            \dot{\hat{\theta}}_{m} &= -\Gamma_m \varphi(x) B (x)^\top M\left(\gamma(1),\hat{\theta}_{em}\right)\gamma_s(1) \\
             \dot{\hat{\theta}}_{em} &= - \Gamma_{em} \varrho(x) M\left(\gamma(1),\hat{\theta}_{em}\right)\gamma_s(1)
        \end{aligned}
    \end{equation}
    where $\mathbbm{1}$ is the indicator vector, $(\cdot)_m$ and $(\cdot)_{em}$ are elements related to the matched or extended matched case, respectively.
\end{remark}

\subsection{Deadzone \& Parameter Bounds}
Several modifications have been proposed in the literature to improve the robustness of adaptive controllers to external disturbances and sensor noise.
Adding a deadzone and leveraging prior knowledge of parameter bounds are two effective means of improving robustness.
\cref{proposition:deadzone} and \cref{proposition:param_bounds} show how a deadzone and parameter bounds can be immediately added to the adaptive controller in \cref{eq:comb_controller}.
Other modifications are also possible but not presented here for brevity.

\begin{proposition}
\label{proposition:deadzone}
Let $\Phi$ denote the size of the deadzone.
The closed-loop system with the modified adaptive controller 
\begin{equation}
    \begin{aligned}
        u &= u_{ccm} + \mathbbm{1} \sum_{i=1}^p \dot{\hat{\theta}}_{em,i} \int \limits_0^1\left[ \frac{\partial \varrho_i\left(\gamma(s)\right)}{\partial x_1} \cdots 0 \right] \gamma_s(s)  ds + \varphi(x)^\top \hat{\theta}_m \\
        \dot{\hat{\theta}}_{m} &= \begin{cases}-\Gamma_m \varphi(x) B (x)^\top M\left(\gamma(1),\hat{\theta}_{em}\right)\gamma_s(1) & |\gamma_s(1)| > \Phi \\ 0 & |\gamma_s(1)| \leq \Phi
        \end{cases} \\
         \dot{\hat{\theta}}_{em} &= \begin{cases} - \Gamma_{em} \varrho(x)  M\left(\gamma(1),\hat{\theta}_{em}\right)\gamma_s(1) & \hspace{1.33cm} |\gamma_s(1)| > \Phi \\ 0 & \hspace{1.33cm} |\gamma_s(1)| \leq \Phi 
         \end{cases}
    \end{aligned}
\end{equation}
asymptotically converges to the deadzone $\Phi$.
\end{proposition}

\begin{proof}
See \nameref{sec:appendix}
\end{proof}

\begin{proposition}
\label{proposition:param_bounds}
Let each parameter $\theta_i$ belongs to the closed convex set $\Upsilon_{\theta_i} = \left\{\theta_i~|~\theta^-_i \leq \theta_i \leq \theta^+_i\right\}$.
The closed-loop system with the modified adaptation law
\begin{equation}
\label{eq:proj}
    \dot{\hat{\theta}}_i = \emph{Proj}_{\hat{\theta}_i}\left(\bullet_i\right) = \begin{cases} 0 \hspace{0.6cm} \hat{\theta}_i = \theta^+_i ~\wedge~ \bullet_i > 0 \\  0 \hspace{0.6cm} \hat{\theta}_i = \theta^-_i ~\wedge~ \bullet_i < 0 \\ \bullet_i \hspace{0.47cm} \emph{Otherwise} \end{cases}
\end{equation}
is asymptotically stable where $\bullet_i$ is the unmodified adaptation law in \cref{eq:comb_controller}.
\end{proposition}

\begin{proof}
See \nameref{sec:appendix}
\end{proof}

\subsection{Offline/Online Computation}
\label{sub:computation}
The proposed adaptive controller has both an offline and online optimization component.
The offline component involves synthesizing the (parameter-dependent) contraction metric.
If the dynamics \textit{orthogonal} to the input matrix are polynomial, then the metric synthesis can be formulated as as sum-of-square (SOS) optimization problem \cite{manchester2017control} and easily solved using existing software packages; gridding can be used if the dynamics are not polynomial.
The SOS CCM formulation is convex which eliminates the complexities associated with SOS CLFs.
The online implementation entails solving a nonlinear optimization problem for a minimizing geodesic at each control cycle.
The necessity for computing a minimal geodesic is not a consequence of the proposed adaptation law but rather constructing a distance-like function i.e., Riemannian energy \emph{differentially}.  
Fortunately, finding a geodesic via optimization is far less computationally expensive than nonlinear MPC due to the absence of dynamic constraints and, under reasonable conditions, a solution is guaranteed to exist.
A continuous-time dynamic realization has been recently proposed \cite{wang2019continuous} that avoids using online optimization to find geodesics but requires an accurate model of the system dynamics. 
Adding adaptation to this framework would be an interesting extension as it would improve robustness to model uncertainty. 
Similar to \cite{leung2017nonlinear}, this work used Chebychev Pseudospectral method and Clenshaw-Curtis quadrature scheme to calculate minimizing geodesics at each time step.

\section{Illustrative Example}
\label{sec:results}

Consider the nonlinear system that has both matched and extended matched parametric uncertainty
\begin{equation}
\left[ \begin{array}{c} \dot{x}_1 \\ \dot{x}_2 \\ \dot{x}_3 \end{array} \right] = \left[ \begin{array}{c}  x_3 \\ x^2_1 - x_2  \\ \mathrm{tanh}(x_2) \end{array} \right] - \theta_1\left[\begin{array}{c}x_1 \\0\\0\end{array}\right]  + \left[ \begin{array}{c} 0 \\ 0 \\ 1 \end{array} \right] \left[u - \theta_2 x_3 - \theta_3 x_1^2 \right],
\label{eq:unc_sys}
\end{equation}
where $\theta_i$ for $i=1,\dots,3$ are unknown parameters.
System \cref{eq:unc_sys} is not feedback linearizable (controllability matrix drops rank at the origin), cannot be put into strict-feedback form, and has non-polynomial entries.
Formulating the constructive dual CCM condition as a sum-of-squares optimization with the YALMIP toolbox \cite{Lofberg2004}, the parameter-dependent dual contraction metric takes the form
\begin{equation}
W\left(x_1,{\theta}_1\right) = \left[ \begin{array}{c c c} 1.42 & 0 & 1.42(\theta_1-1) \\ 
0 & 6.21 & -2.85x_1 \\ 
1.42(\theta_1-1) & -2.85x_1 & 1.42\theta_1^2 - 2.84\theta_1 + 1.30x_1^2 + 5.79 \end{array} \right] ,
\label{eq:m_ex}
\end{equation}
which is non-flat indicating \cref{eq:unc_sys} cannot be stabilized with a quadratic control Lyapunov function.
Let $\hat{\theta}_m$ and $\hat{\theta}_{em}$ denote the estimated matched and extended matched parameters, respectively.
Then, given a minimizing geodesic $\gamma(s)$ that connects the current and desired state, the CCM controller $u_{ccm}$ for the parameter estimate $\hat{\theta}_{em}$ is found by solving the quadratic program
\begin{equation}
\begin{aligned}
u_{ccm} = ~ & \underset{u}{\mathrm{\textbf{argmin}}} ~ u^\top u \\
& \textbf{subject to} \\
&  \gamma_s(1)^\top M\left(\gamma(1),\hat{\theta}_{em}\right)\{f(x) - \varrho(x)^\top \hat{\theta}_{em} +B(x)\left[u_d+u\right]\} \\
& \hspace{6cm} -\gamma_s(0)^\top M\left(\gamma(0),\hat{\theta}_{em}\right)\dot{x}_d 
\leq -  \lambda E(x,x_d,\hat{\theta}_{em})
\end{aligned}
\label{eq:ccm_qp}
\end{equation}
where $E(x,x_d,\hat{\theta}_{em})$ is the Riemannian energy and calculated through quadrature.
The controller $u_{ccm}$ is commonly referred to as the pointwise min-norm controller \cite{primbs2000receding}. 
The adaptive CCM controller takes the final form
\begin{equation}
\label{eq:accm}
    \begin{aligned}
        u &= u_{ccm} + \dot{\hat{\theta}}_1 \left(x_1-x_{1d}\right) + \hat{\theta}_2 x_3 + \hat{\theta}_3 x_1^2 \\
        \dot{\hat{\theta}}_1 &= -\Gamma_1 \left[ x_1~0 ~ 0 \right]  M\left(\gamma(1),\hat{\theta}_1\right)\gamma_s(1) \\
        \left[\begin{array}{c}
        \hat{\theta}_2 \\
        \hat{\theta}_3
        \end{array} \right] &= - \Gamma_2 \left[ \begin{array}{c} x_3 \\ x_1^2 \end{array} \right] \left[ 0 ~ 0 ~ 1\right] M\left(\gamma(1),\hat{\theta}_1\right)\gamma_s(1),
    \end{aligned}
\end{equation}
where $\Gamma_1$ and $\Gamma_2$ are the adaptation rates for the extended matched and matched uncertainties, respectively.

\begin{figure}[t!]
\centering 
  \subfloat[State trajectories with baseline controller.]{\includegraphics[width=0.3\textwidth]{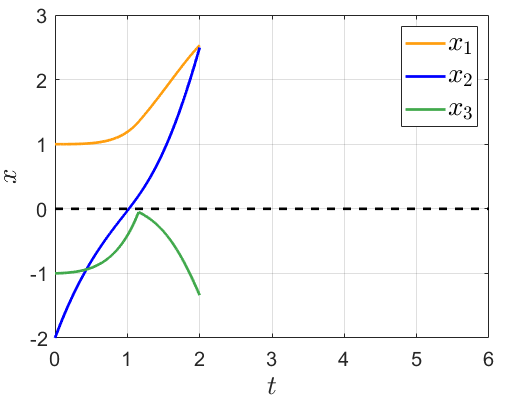}\label{fig:ccm}} 
  \hspace{0.6em}
  \subfloat[State trajectories with adaptive controller.]{\includegraphics[width=0.3\textwidth]{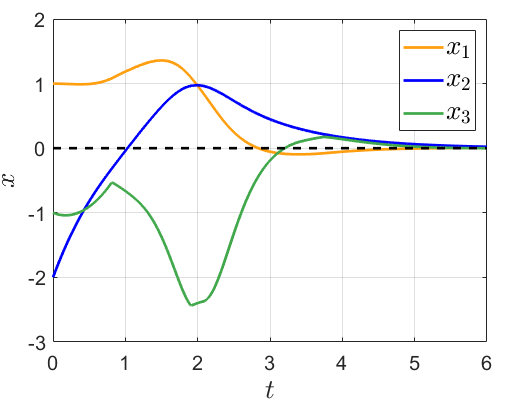}\label{fig:accm}}
  \hspace{0.5em}
  \subfloat[Parameter estimates.]{\includegraphics[width=0.32\textwidth]{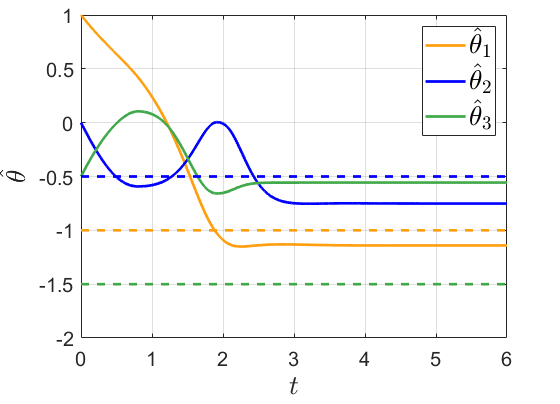}\label{fig:theta}}
  \caption{State trajectories and parameter estimates with the proposed adaptive controller. (a): Closed-loop system without adaptation becomes unstable due to the matched uncertainty. (b): Closed-loop system with adaptation successfully converges to origin with matched and extended matched uncertainty. (c): Evolution of parameter estimates as closed-loop system converges to origin.} 
    \label{fig:accm_results}
    \vskip -0.2in
\end{figure}

The adaptive controller in \cref{eq:accm} was tested on system \cref{eq:unc_sys} with true parameter values of $\theta = \left[ -1~-0.5~-1.5 \right]^\top$ and initial parameter estimates $\hat{\theta}_0 = \left[ 1~0~-0.5 \right]^\top$.
Figure~\ref{fig:ccm} shows the time trace for each state (solid lines) and setpoint (dashed line) when the parameter-dependent contraction metric pointwise min-norm controller (no adaptation) is applied.
At each time-step a minimizing geodesic is calculated through nonlinear optimization as discussed in \cref{sub:computation}.
While the min-norm controller is guaranteed to stabilize the uncertain system with extended matched uncertainty, the structure of the matched uncertainty causes the closed-loop system to go unstable; the simulation had to be terminated at $t=2s$ as a result.
\cref{fig:accm} shows the time trace for each state (sold lines) and setpoint (dashed line) with the adaptive min-norm controller.
Each state converges to their respective desired values further proving the proposed adaptive controller can stabilize \cref{eq:unc_sys} with both matched and extended matched uncertainty.
The parameter estimates (solid lines) and true values (dashed lines) are shown in \cref{fig:theta} for reference.

\begin{figure}[t!]
\centering 
  \subfloat[State trajectories with modified adaptive controller leveraging parameter bounds.]{\includegraphics[width=0.32\textwidth]{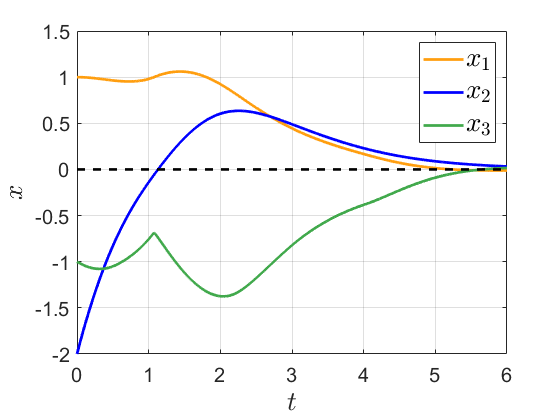}\label{fig:accm_bounds}}
  \hspace{2em}
  \subfloat[Parameter estimates with parameter bounds.]{\includegraphics[width=0.32\textwidth]{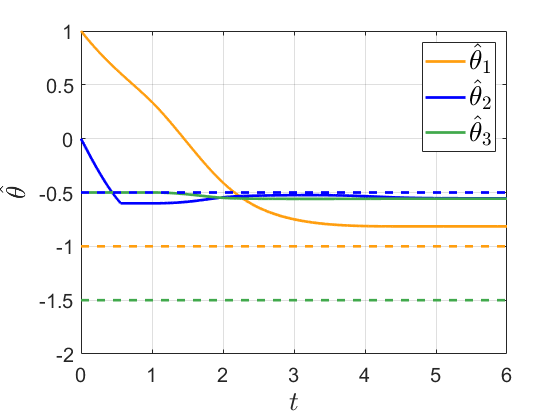}\label{fig:theta_bounds}}
  \caption{State trajectories and parameter estimates with the modified adaptive controller when parameter bounds are known (\cref{proposition:param_bounds}). (a): Closed-loop system successfully converges to origin with matched and extended matched uncertainty with modified adaptive controller leveraging parameter bounds. (b): Evolution of parameter estimates as closed-loop system converges to origin.} 
    \label{fig:accm_bounds_results}
    \vskip -0.2in
\end{figure}

Better transient performance can be obtained if parameter bounds are incorporated into the adaptation law.
\cref{fig:accm_bounds} shows the time trace for each state (sold lines) as they converge to the desired setpoint (dashed line) with the modified adaptive controller from \cref{proposition:param_bounds}.
Comparing to \cref{fig:accm}, the peak transient error at $t=2$s is reduced by 33.3\%. 
Moreover, the parameter estimates, shown in \cref{fig:theta_bounds}, exhibit less overshoot as the system converge to the desired setpoint. 

\section{Discussion}
This work used contraction analysis to synthesize an adaptive controller for nonlinear systems without constructing an explicit control Lyapunov function.
Consequently, the proposed controller can be applied to any stabilizable nonlinear system that is path integrable; far less restrictive conditions than those needed for feedback linearization and backstepping.
The main drawback of using a differential control Lyapunov function is the need to compute online minimizing geodesics and performing quadrature.
While computing minimizing geodesics involves solving a nonlinear optimization problem, the computational complexity is less than that of nonlinear model predictive control and a solution is guaranteed to exist under suitable conditions.
The matched uncertainty controller in \cref{sub:matched} is a generalization of that presented in \cite{slotine2003modular} which relied on constructing an explicit change of variables $z = \Theta(x)$.
Moreover, the extended matched controller in \cref{sub:exMatched} is a generalization of those derived using feedback linearization and backstepping.
Avoiding an explicit coordinate transformation is the primary reason why this work can be applied to a broader class of nonlinear systems. 

The adaptive controller developed in \cref{sec:accm} augments a nominal controller with adaptive feedback terms.
While the pointwise min-norm controller was used in this work, other control policies can be used so long as they are known to be contracting in some metric.
For instance, if a learned policy $\pi(x,\theta)$ is known to be contracting in metric $M_{\pi}(x,\theta)$ for all possible $\theta$, then adaptation can be immediately added to the learned policy by
 \begin{equation}
    \begin{aligned}
        \pi_a &= \pi(x,\hat{\theta}) + \mathbbm{1} \sum_{i=1}^p \dot{\hat{\theta}}_{em,i} \int \limits_0^1\left[ \frac{\partial \varrho_i\left(\gamma(s)\right)}{\partial x_1} \cdots 0 \right] \gamma_s(s)  ds + \varphi(x)^\top \hat{\theta}_m \\
        \dot{\hat{\theta}}_{m} &= - \Gamma_m \varphi(x) B (x)^\top M_\pi\left(\gamma(1),\hat{\theta}_{em}\right)\gamma_s(1) \\
         \dot{\hat{\theta}}_{em} &= - \Gamma_u \varrho(x)  M_\pi\left(\gamma(1),\hat{\theta}_{em}\right)\gamma_s(1)
    \end{aligned}
\end{equation}
where $\hat{\theta}_m$ and $\hat{\theta}_{em}$ are the estimated matched and extended matched parameters, respectively.
Therefore, \textit{any} robust feedback policy, learned or otherwise, that is known to be contracting in some metric can be combined with the proposed adaptive controller to achieve zero tracking error for any feasible desired trajectory. 
Augmenting learned feedback policies with adaptation will be critical for future applications (especially robotics) as the desire to achieve complex tasks will necessarily require high-performance control.
Learning contracting feedback policies for high-dimensional nonlinear systems and addressing the more general extended uncertainty case are both future work.

\section*{Appendix}
\label{sec:appendix}

\begin{proof}[Proof of Lemma~\ref{lemma:matched}]
    Assume there exists a uniformly bounded contraction metric $M(x)$ for the differential dynamics of the \textit{nominal} system $\dot{\delta}_x = \bar{A}(x,u)\delta_x + B(x)\delta_u$, where $\bar{A}(x,u) = \frac{\partial f}{\partial x} + \sum_{i=1}^m \frac{\partial b_i}{\partial x}u_i$, that satisfies the following implication
    \begin{equation*}
        \delta_x^\top M B = 0 \implies \delta_x^\top \left( \bar{A}^\top M + M \bar{A} + \dot{M} + 2 \lambda M \right) \delta_x \leq 0.
    \end{equation*}
    As proposed in \cite{manchester2017control}, through a change of variables $W(x) = M(x)^{-1}$ and $B_{\perp}(x) = M(x)\delta_x$ where $B_{\perp}(x)^\top B(x) = 0$, the constructive condition for $W(x)$ (the dual metric) is given by
    \begin{equation}
        B_{\perp}^\top \left(  W \bar{A}^\top + \bar{A} W - \dot{W} + 2 \lambda W \right) B_{\perp} \preceq 0.
    \label{eq:dualCCMCond}
    \end{equation}
    Imposing each $b_i$ for $i=1,\dots,m$ forms a Killing vector field for $W(x)$ , then -$\sum_{j=1}^n \frac{\partial W}{\partial x_j} b_{i,j} + \frac{\partial b_i}{\partial x}^\top W + W \frac{\partial b_i}{\partial x} = 0$.
    So \cref{eq:dualCCMCond} can be further simplified to
    \begin{equation*}
        B_{\perp}^\top \left(  W \frac{\partial f}{\partial x}^\top + \frac{\partial f}{\partial x} W - \dot{W} + 2 \lambda W \right) B_{\perp} \preceq 0.
    \end{equation*}
    Now consider the differential dynamics of the \textit{actual} system 
    \begin{equation*}
        A(x,u) = \frac{\partial f}{\partial x} + \sum_{i=1}^m \frac{\partial b_i(x)}{\partial x}\left[u_i-\varphi_i(x)^\top \theta\right] + \sum_{i=1}^m b_i(x) \frac{\partial \varphi_i(x)}{\partial x}^\top \theta_i.
    \end{equation*}
    Since each $b_i(x)$ forms a Killing vector field for $W(x)$ and satisfies $b_{i_\perp}(x)^\top b_i(x) = 0$, then the second and third term of $A(x,u)$ vanish when substituted into the dual CCM condition \cref{eq:dualCCMCond}.
    Therefore the constructive conditions for the true systems are identical to that of the nominal system.
\end{proof}

\begin{proof}[Proof of Theorem~\ref{thm:matched}]
    From Assumption~\ref{assumption:ccmExists}, there exists a contraction metric $M(x)$ that satisfies the stronger CCM condition for $\dot{x} = f(x)+ B(x)u$.
    Since the uncertainty satisfies Assumption~\ref{assumption:match}, $M(x)$ is also a valid contraction metric for the uncertain system by \cref{lemma:matched}.
    Let $u_{ccm}$ denote the CCM controller for the nominal system.
    Taking the first variation of the Riemannian energy with the actual dynamics,
    \begin{equation*}
        \dot{E}(x,x_d) = 2 \gamma_s(1)^\top M(\gamma(1)) \{f(x)  + B(x)\left[u - \varphi(x)^\top \theta \right]\} - 2 \gamma_s(0)^\top M(\gamma(0)) \{f(x_d) + B(x_d)u_d\}.
    \end{equation*}
    Letting $u = u_{ccm} + \varphi(x)^\top \hat{\theta}$
    \begin{equation*}
    \begin{aligned}
        \dot{E}(x,x_d) = 2 \gamma_s(1)^\top M(\gamma(1)) \{f(x)  + B(x)u_{ccm}\} - 2 \gamma_s(0)^\top & M(\gamma(0))  \{f(x_d) + B(x_d)u_d\} \\
         & + 2 \gamma_s(1)^\top M(\gamma(1)) B(x) \varphi(x)^\top \tilde{\theta},
    \end{aligned}
    \end{equation*}
    where $\tilde{\theta} := \hat{\theta}-\theta$. By Assumption~\ref{assumption:ccmExists}, the following bound can be imposed on the first variation of the Riemannian energy
    \begin{equation*}
        \dot{E}(x,x_d) \leq - 2 \lambda E(x,x_d) + 2 \gamma_s(1)^\top M(\gamma(1)) B(x) \varphi(x)^\top \tilde{\theta}.
    \end{equation*}
    Let $L(x,x_d)$ be the output of the first-order filter
    \begin{equation}
    \label{eq:L_def}
        \dot{L}(x,x_d) + 2 \lambda L(x,x_d) = 2 \gamma_s(1)^\top M(\gamma(1)) B(x) \varphi(x)^\top \tilde{\theta}.
    \end{equation}
    Selecting the adaptation law
    \begin{equation*}
        \dot{\hat{\theta}} = -\Gamma \varphi(x) B (x)^\top M(\gamma(1))\gamma_s(1),
    \end{equation*}
    one can construct the following system
    \begin{equation*}
        \left[ \begin{array}{cc}
            \frac{1}{2} & 0  \\
             0 & \Gamma^{-1}
        \end{array} \right]
        \left( \begin{array}{c}
             \dot{L}(x,x_d) \\
             \dot{\tilde{\theta}}
        \end{array}\right) = \left[ \begin{array}{cc}
             -\lambda & \star  \\
            - \star^\top & 0
        \end{array} \right] \left( \begin{array}{c}
             L(x,x_d) \\
             \tilde{\theta} \end{array}  \right),
    \end{equation*}
    where $\star := \gamma_s(1)^\top M(\gamma(1)) B(x) \varphi(x)^\top$.
    Consider the virtual system
    \begin{equation*}
        \left[ \begin{array}{cc}
            \frac{1}{2} & 0  \\
             0 & \Gamma^{-1}
        \end{array} \right] \left( \begin{array}{c}
             \dot{y}_1 \\
             \dot{y}_2
        \end{array} \right) = \left[ \begin{array}{cc}
             -\lambda & \star  \\
            - \star^\top & 0
        \end{array} \right] \left( \begin{array}{c}
             y_1 \\
             y_2
        \end{array} \right),
    \end{equation*}
    which has two particular solutions, mainly $y_1 = 0$, $y_2 = 0$  and $y_1 = L(x,x_d)$, $y_2 = \tilde{\theta}$.
    Consider the Lyapunov-like function
    \begin{equation*}
        V = \left( \begin{array}{c}
             y_1 \\
             y_2
        \end{array} \right)^\top \left[ \begin{array}{cc}
            \frac{1}{2} & 0  \\
             0 & \Gamma^{-1}
        \end{array} \right] \left( \begin{array}{c}
             y_1 \\
             y_2
        \end{array} \right),
    \end{equation*}
    yielding $\dot{V} = -2 \lambda y_1 ^ 2 \leq 0$ , which implies that both $y_1$ and $y_2$ are bounded.
    Moreover, $\ddot{V} = - 4 \lambda y_1 \dot{y_1}$ is also bounded, because all terms on the right hand side of the first-order filter \cref{eq:L_def} are bounded.
    First, note that $E(x,x_d) \geq 0$ and by the Comparison Lemma \cite{khalil2002nonlinear} $0 \leq E(x,x_d) \leq L(x,x_d)$ for all $t$ so $E(x,x_d)$ must be bounded since $L(x,x_d)$ is bounded.
    Then, $\tilde{\theta}$ is bounded since $\dot{V} \leq 0$, if $B(x)$ and $\varphi(x)$ have bounded output for bounded input then they are bounded since $E(x,x_d)$ is bounded so $x$ is bounded for bounded $x_d$, and $M$ is bounded by definition.
    Since geodesics have constant speed then $E(x,x_d) = \int_0^1 \left<\gamma_s(s),\gamma_s(s)\right> ds = \left<\gamma_s,\gamma_s\right>$ and since $E(x,x_d)$ is bounded and $M$ is positive definite then $\gamma_s(s)$ must also be bounded for all $s\in[0~1]$.
    The conditions for Barbalat's Lemma are then satisfied so $L(x,x_d) \rightarrow 0$.
    By the Comparison Lemma $E(x,x_d) \rightarrow 0$ and $x\rightarrow x_d$ asymptotically as desired.

\end{proof}

\begin{proof}[Proof of \cref{lemma:ccmRobust}]
    Let $u = u_{ccm} + \varphi(x)^\top \hat{\theta} + u_{R}$.
    From Assumption~\ref{assumption:ccmExists}, the contracting baseline controller $u_{ccm}$ ensures the first variation of the Riemannian energy satisfies the inequality
    \begin{equation*}
        \dot{E}(x,x_d) \leq - 2 \lambda E(x,x_d) + 2 \gamma_s(1)^\top M (\gamma(1)) B(x) \left[ u_R + \varphi(x)^\top \tilde{\theta} \right], 
    \end{equation*}
    or 
    \begin{equation}
    \label{eq:E_dot_r}
        \dot{E}(x,x_d) \leq - 2 \lambda E(x,x_d) + 2 \sum_{i=1}^m \gamma_s(1)^\top M (\gamma(1)) b_i(x) \left[ u_{R,i} + \varphi_i(x)^\top \tilde{\theta} \right], 
    \end{equation}
    where $\tilde{\theta}:=\hat{\theta}-\theta$.
    Let 
    \begin{equation*}
        u_{R,i} =  - \kappa ~ b_i(x)^\top  M(\gamma(1))\gamma_s(1) \|\varphi_i(x)\|^2,
    \end{equation*}
    where $\kappa > 0$ is a design parameter.
    Then, \cref{eq:E_dot_r} becomes
    \begin{equation*}
    \begin{aligned}
        \dot{E}(x,x_d) \leq -2 \lambda E(x,x_d) - 2 & \sum_{i=1}^m \left[ \kappa \|\gamma_s(1)^\top M(\gamma(1)) b_i(x) \|^2 \|\varphi_i(x)\|^2 \right. \\ 
        & \hspace{3.5cm} \left. - \gamma_s(1)^\top M(\gamma(1)) b_i(x)  \varphi_i(x)^\top \tilde{\theta} \right].
    \end{aligned}
    \end{equation*}
    Completing the square,
    \begin{equation*}
    \begin{aligned}
        \dot{E}(x,x_d) & \leq -2 \lambda E(x,x_d) - 2 \kappa \sum_{i=1}^m \left[ \left|\left| \gamma_s(1)^\top M(\gamma(1)) b_i(x) \varphi_i(x)  -  \frac{1}{4\kappa} \tilde{\theta} \right|\right|^2 \right]  + \frac{m}{2 \kappa} \left|\left|\tilde{\theta}\right|\right|^2 \\
        & \leq -2 \lambda E(x,x_d) + \frac{m}{2 \kappa} \left|\left|\tilde{\theta}\right|\right|^2
    \end{aligned}
    \end{equation*}
    From Assumption~\ref{assumption:paramBounds} and letting $K:=\frac{m}{2\kappa}$, one obtains \cref{eq:E_bound} with \cref{eq:uccmRobust}.
\end{proof}

\begin{proof}[Proof of \cref{thm:accm_transient}]
Taking the first variation of the Riemannian energy and applying control law \cref{eq:uccmRobust},
\begin{equation*}
\begin{aligned}
    \dot{E}(x,x_d) &\leq - 2 \lambda E(x,x_d) + 2  \gamma_s(1)^\top M(\gamma(1)) B(x) \varphi(x)^\top \tilde{\theta} \\
    & \hspace{5cm} - 2 \kappa \sum_{i=1}^m   \|\gamma_s(1)^\top  M(\gamma(1)) b_i(x) \|^2 \|\varphi_i(x)\|^2\\
    &\leq - 2 \lambda E(x,x_d) + 2  \gamma_s(1)^\top M(\gamma(1)) B(x) \varphi(x)^\top \tilde{\theta},
\end{aligned}
\end{equation*}
where the last term in the first inequality is always negative semi-definite and bounded.
Picking the same adaptation law in \cref{thm:matched}, asymptotic stability is proven through the same process used in \cref{thm:matched} where a Lyapunov-like function for a virtual system is shown to satisfy Barbalat's lemma. 
Moreover, since the conditions for \cref{lemma:ccmRobust} hold, the transients of the Riemannian energy, and hence the tracking error, is exponentially bounded.
\end{proof}

\begin{proof}[Proof of \cref{lemma:exMatched}]
    Restricting the uncertainty to be linear in unknown parameters and one derivative away from the input imposes structure on the parameter-dependent metric $M(x,\hat{\theta})$.
    Let $\varrho_i(x)$ denote the $i^{\text{th}}$ row vector of $\varrho(x)$.
    The elements of metric $M$ will then be at most quadratic in  $\hat{\theta}_i$ and basis function $\frac{\partial \varrho_i}{\partial x_j}$ for all $j$, with the quadratic terms located along the main diagonal of $M$ excluding the last diagonal entry.
    The off-diagonal elements are at most linear in $\theta_i$ and $\frac{\partial \varrho_i}{\partial x_j}$ for all $j$.
    
    Now consider the differential change of coordinates $\delta_z = \Theta(x,\theta) \delta_x$ where $\Theta \succ 0$ is a lower triangular matrix such that $M(x,\theta) = \Theta(x,\theta)^\top \Theta(x,\theta)$.
    Since $M(x,\theta)$ is positive definite in both $x$ and $\theta$ then $\Theta(x,\theta)$ is guaranteed to exist and be unique. The expression \cref{eq:ex_rate_cond} is equivalent to showing
    \begin{equation}
    \label{eq:ex_rate_theta}
        \frac{\partial \Theta}{\partial \hat{\theta}_i} = - \Theta b_k \left[ \frac{\partial \varrho_i}{\partial x_1} \cdots 0 \right],
    \end{equation}
    where $b_k(x)$ is a column vector of $B(x)$ such that $\varrho_i(x)^\top \theta_i \in \text{span}\{ad_f b_k\}$.
    Leveraging the structure of $M(x,\theta)$, the elements along the principle diagonal and below of $\Theta(x,\theta)$ are at most linear in $\theta_i$ and $\frac{\partial \varrho_i}{\partial x_j}$ for all $j$.
    Hence, the left hand side of \cref{eq:ex_rate_theta} will by a matrix with elements linear in $\frac{\partial \varrho_i}{\partial x_j}$ for all $j < n$.
    Without loss of generality let the last diagonal element of $\Theta$ be 1 through appropriate scaling.
    Since $\Theta(x,\theta)$ is lower triangular, then 
    \begin{equation}
    \label{eq:ex_rate_theta_lhs}
        \Theta b_k \left[ \frac{\partial \varrho_i}{\partial x_1} \cdots 0 \right] = \left[\begin{array}{c}0 \\ \vdots\\  1\end{array}\right] \left[ \frac{\partial \varrho_i}{\partial x_1} \cdots 0 \right]
    \end{equation}
    which results in a matrix identical $\frac{\partial \Theta}{\partial \hat{\theta}_i}$ but of opposite sign due to the definition of the uncertainty in \cref{eq:exMatchedDyn}.
    Negating \cref{eq:ex_rate_theta_lhs} leads to equality and hence \cref{eq:ex_rate_cond}.
\end{proof}

\begin{proof}[Proof of \cref{thm:exMatched}]
    From Assumption~\ref{assumption:exCCMexists}, there exists a parameter-dependent contraction metric $M(x,\theta)$ that satisfies the CCM condition for \textit{all} values of $\theta$.
    Replacing $\theta$ with $\hat{\theta}$ and using the same differential CLF $\delta_V = \delta_x^\top M(x,\hat{\theta})\delta_x$, stability is achieved if
    \begin{equation}
        \dot{\delta}_V  = \delta_x^\top \left(A^\top M + MA + \frac{\partial M}{\partial t} + \partial_f M \right)\delta_x + \sum_{i=1}^p\delta_x^\top \dot{\hat{\theta}}_i \frac{\partial M}{\partial \hat{\theta}_i} \delta_x + 2 \delta_x^\top MB\delta_u \leq -2 \lambda \delta_x^\top M \delta_x.
        \tag{\ref{eq:dclf_dot_adapt}}
    \end{equation}
    Applying \cref{lemma:exMatched}, \cref{eq:dclf_dot_adapt} becomes
    \begin{equation*}
    \begin{aligned}
        \dot{\delta}_V  = & \delta_x^\top \left(A^\top M + MA + \frac{\partial M}{\partial t} + \partial_f M \right)\delta_x \\
        & \hspace{1cm} -  2\sum_{i=1}^p\delta_x^\top Mb_k \dot{\hat{\theta}}_i \left[ \frac{\partial \varrho_i}{\partial x_1} \cdots 0 \right] \delta_x + 2 \delta_x^\top MB\delta_u \leq -2 \lambda \delta_x^\top M \delta_x,
    \end{aligned}
    \end{equation*}
    which can be further simplified to
    \begin{equation*}
    \begin{aligned}
        \dot{\delta}_V  = & \delta_x^\top \left(A^\top M + MA + \frac{\partial M}{\partial t} + \partial_f M \right)\delta_x \\
        & \hspace{1cm} + 2\delta_x^\top M B \left[-\mathbbm{1} \sum_{i=1}^p \dot{\hat{\theta}}_i  \left[ \frac{\partial \varrho_i}{\partial x_1} \cdots 0 \right] \delta_x + \delta_u \right] \leq -2 \lambda \delta_x^\top M \delta_x,
    \end{aligned}
    \end{equation*}
    where $B(x)\mathbbm{1} = b_k(x)$ so $\varrho_i(x)^\top \theta_i \in \text{span}\{ad_fb_k\}$.
    Picking $\delta_u = \delta_{u_{ccm}} + \mathbbm{1} \sum_{i=1}^p \dot{\hat{\theta}}_i \left[ \frac{\partial \varrho_i}{\partial x_1} \cdots 0 \right] \delta_x$, the original differential CLF stability condition is obtained
    \begin{equation*}
        \dot{\delta}_V  = \delta_x^\top \left(A^\top M + MA + \frac{\partial M}{\partial t} + \partial_f M \right)\delta_x + 2\delta_x^\top M B \delta_{u_{ccm}} \leq -2 \lambda \delta_x^\top M \delta_x.
    \end{equation*}
    Following the same procedure of that taken in \cref{thm:matched},
    taking the first variation of the Riemannian energy with the actual system,
    \begin{equation}
    \begin{aligned}
        \dot{E}(x,x_d,\hat{\theta}) = \frac{\partial E}{\partial \hat{\theta}}\dot{\hat{\theta}} + 2 \gamma_s(1)^\top   M\left(\gamma(1),\hat{\theta}\right) & \{f(x) - \varrho(x)^\top \theta + B(x)u\}  \\ 
        &- 2 \gamma_s(0)^\top M\left(\gamma(0),\hat{\theta}\right) \{f(x_d) - \varrho(x)^\top \theta + B(x_d)u_d\}
        \label{eq:diff_clf_ex}
    \end{aligned}
    \end{equation}
    where the term $\frac{\partial E}{\partial \hat{\theta}}\dot{\hat{\theta}}$ results from the parameter-dependent contraction metric.
    Letting 
    \begin{equation*}
        u=u_{ccm} + \mathbbm{1} \sum_{i=1}^p \dot{\hat{\theta}}_i \int \limits_0^1 \left[ \frac{\partial \varrho_i\left(\gamma(s)\right)}{\partial x_1} \cdots 0 \right] \gamma_s(s) ds
    \end{equation*}
    and noting that, by construction, the second term in the controller cancels the term $\frac{\partial E}{\partial \hat{\theta}}\dot{\hat{\theta}}$.
    Applying the definition if $\tilde{\theta}$, \cref{eq:diff_clf_ex} becomes
    \begin{equation*}
        \begin{aligned}
            \dot{E}(x,x_d,\hat{\theta}) = 2 \gamma_s(s)^\top M\left(\gamma(1),\hat{\theta}\right) & \varrho(x)^\top   \tilde{\theta} +  2  \gamma_s(1)^\top   M\left(\gamma(1),\hat{\theta}\right) \{f(x) - \varrho(x)^\top \hat{\theta} + B(x)u_{ccm}\} \\ & - 2 \gamma_s(0)^\top M\left(\gamma(0),\hat{\theta}\right)  \{f(x_d) - \varrho(x)^\top \theta + B(x_d)u_d\}.
        \end{aligned}
    \end{equation*}
    By Assumption~\ref{assumption:exCCMexists}, the first variation of the Riemannian energy satisfies
    \begin{equation*}
        \dot{E}(x,x_d,\hat{\theta}) \leq  - 2 \lambda E(x,x_d,\hat{\theta})  + 2 \gamma_s(s)^\top M\left(\gamma(1),\hat{\theta}\right) \varrho(x)^\top \tilde{\theta}.
    \end{equation*}
    Again letting $L(x,x_d,\hat{\theta})$ be output of the first order filter 
    \begin{equation}
    \label{eq:L_def_exmatched}
        \dot{L}(x,x_d,\hat{\theta}) + 2 \lambda L(x,x_d,\hat{\theta}) = 2 \gamma_s(s)^\top M\left(\gamma(1),\hat{\theta}\right) \varrho(x)^\top \tilde{\theta},
    \end{equation}
    and selecting the adaptation law
    \begin{equation*}
        \dot{\hat{\theta}} = - \Gamma \varrho(x)^\top  M\left(\gamma(1),\hat{\theta}\right)\gamma_s(1),
    \end{equation*}
    one can construct the virtual system 
    \begin{equation*}
        \left[ \begin{array}{cc}
            \frac{1}{2} & 0  \\
             0 & \Gamma^{-1}
        \end{array} \right] \left( \begin{array}{c}
             \dot{y}_1 \\
             \dot{y}_2
        \end{array} \right) = \left[ \begin{array}{cc}
             -\lambda & \star  \\
            - \star^\top & 0
        \end{array} \right] \left( \begin{array}{c}
             y_1 \\
             y_2
        \end{array} \right),
    \end{equation*}
    where $\star := \gamma_s(1)^\top M(\gamma(1))\varrho(x)^\top$.
    The virtual system has two particular solutions, mainly $y_1 = 0$, $y_2 = 0$  and $y_1 = L(x,x_d)$, $y_2 = \tilde{\theta}$.
    Consider the Lyapunov-like function
    \begin{equation*}
        V = \left( \begin{array}{c}
             y_1 \\
             y_2
        \end{array} \right)^\top \left[ \begin{array}{cc}
            \frac{1}{2} & 0  \\
             0 & \Gamma^{-1}
        \end{array} \right] \left( \begin{array}{c}
             y_1 \\
             y_2
        \end{array} \right),
    \end{equation*}
    yielding $\dot{V} = -2 \lambda y_1 ^ 2 \leq 0$ which implies both $y_1$ and $y_2$ are bounded.
    Moreover, $\ddot{V} = - 4 \lambda y_1 \dot{y_1}$ is also bounded because all terms on the right hand side of the first-order filter \cref{eq:L_def_exmatched} are bounded following similar arguments as those presented in \cref{thm:matched}.
    Since $E(x,x_d,\hat{\theta}) \geq 0$ then by the Comparison Lemma  $0 \leq E(x,x_d,\hat{\theta}) \leq L(x,x_d,\hat{\theta})$ for all $t$.
    As $L(x,x_d,\hat{\theta}) \rightarrow 0$ then $E(x,x_d,\hat{\theta}) \rightarrow 0$ and $x\rightarrow x_d$ asymptotically as desired.
\end{proof}

\begin{proof}[Proof of \cref{proposition:deadzone}]
    Follows immediately from \cref{thm:matched} and \cref{thm:exMatched} when $|\gamma_s(1)| \geq \Phi$.
\end{proof}
    
\begin{proof}[Proof of \cref{proposition:param_bounds}]
    For clarity only the matched uncertainty case will be considered as identical arguments can be made for the extended matched case.
    Using the control law $u = u_{ccm} + \varphi(x)^\top \hat{\theta}$, the first variation of the Riemannian energy satisfies the inequality
    \begin{equation*}
        \dot{E}(x,x_d) \leq -2 \lambda E(x,x_d) + 2 \gamma_s(1)^\top M(\gamma(1))B(x)\varphi(x)^\top \tilde{\theta}.
    \end{equation*}
    \cref{thm:matched} showed that the closed-loop system is asymptotically stable by selecting the appropriate adaptation law.
    Let $\varphi_k(x)$ be the $k^{\text{th}}$ row of $\varphi(x)$.
    If adaptation is temporarily stopped for parameter ${\theta}_{k}$ based on the conditions in \cref{eq:proj}, then
    \begin{equation*}
    \begin{aligned}
    \label{eq:E_dot_bounds}
        \dot{E}(x,x_d) &\leq - 2 \lambda E(x,x_d) + 2 \gamma_s(1)^\top M(\gamma(1))B(x)\varphi_k(x)^\top \tilde{\theta}_k + \sum_{j\neq k}^p 2 \gamma_s(1)^\top M(\gamma(1))B(x)\varphi_j(x)^\top \tilde{\theta}_j  \\
        & \leq - 2 \lambda E(x,x_d) + \sum_{j\neq k}^p 2 \gamma_s(1)^\top M(\gamma(1))B(x)\varphi_j(x)^\top \tilde{\theta}_j.
    \end{aligned}
    \end{equation*}
    The second term to the right of the first inequality is negative semi-definite when the stopping conditions for the projection operator are true, leading to the second inequality.
    Asymptotic stability is shown using the same procedure as in \cref{thm:matched} but with $\dot{\hat{\theta}}_k = 0$.
    Hence, $E(x,x_d) \rightarrow 0$ and $x \rightarrow x_d$ as desired.

\end{proof}

\noindent \textbf{Acknowledgements}  
We thank Sumeet Singh and Stephen Tu for stimulating discussions.


\balance
\bibliographystyle{ieeetr}
\bibliography{ref}

\end{document}